\documentclass[aps,prb,preprint,superscriptaddress]{revtex4-1}
\usepackage[pdftex,colorlinks=true,linkcolor=blue,citecolor=blue,urlcolor=blue]{hyperref}
\usepackage{hyperref}
\usepackage{graphicx}
\usepackage[utf8]{inputenc}
\usepackage{booktabs}
\newcommand{\figuremacro}[3]{
	\begin{figure}[tbp]
		\centering
		\includegraphics[width=#3\textwidth]{#1}
		\caption[]{#2}
		\label{fig:#1}
	\end{figure}
}

\begin{document}

\title{Lattice dynamics of MgSiO$_3$ perovskite (bridgmanite) studied by inelastic x-ray scattering and \textit{ab initio} calculations}

\author{Bj\"orn Wehinger}
\email[]{bjorn.wehinger@unige.ch}
\affiliation{Department of Quantum Matter Physics, University of Geneva, 24, Quai Ernest Ansermet, CH-1211 Gen\`eve, Switzerland}
\affiliation{Laboratory for Neutron Scattering and Imaging, Paul Scherrer Institute, CH-5232
Villigen PSI, Switzerland}

\author{Alexe\"i Bosak}
\affiliation{European Synchrotron Radiation Facility, 71, Avenue des Martyrs, F-38000 Grenoble, France}

\author{Sabrina Nazzareni}
\affiliation{Department of Physics and Geology, University of Perugia, via Pascoli, I-06123 Perugia, Italy}

\author{Daniele Antonangeli}
\affiliation{Institut de Minéralogie, de Physique des Matériaux, et de Cosmochimie (IMPMC), UMR CNRS 7590, Sorbonne Universités – UPMC, Muséum National d’Histoire Naturelle, 75252 Paris, France}

\author{Alessandro Mirone}
\affiliation{European Synchrotron Radiation Facility, 71, Avenue des Martyrs, F-38000 Grenoble, France}

\author{Samrath Lal Chaplot}
\affiliation{Solid State Physics Division, Bhabha Atomic Research Centre, Mumbai 400085, India}

\author{Ranjan Mittal}
\affiliation{Solid State Physics Division, Bhabha Atomic Research Centre, Mumbai 400085, India}

\author{Eiji Ohtani}
\affiliation{Department of Earth Science, Graduate School of Science, Tohoku University, Sendai 980-8578, Japan}

\author{Anton Shatskiy}
\affiliation{V.S. Sobolev Institute of Geology and Mineralogy, Russian Academy of Science, Siberian Branch, Koptyuga pr. 3, Novosibirsk 630090, Russia}

\author{Saxena Surendra}
\affiliation{Center for the Study of Matter at Extreme Conditions. Florida International University, Miami, USA}

\author{Subrata Ghose}
\thanks{Deceased January 21, 2015}
\affiliation{Department of Earth and Space Sciences, University of Washington, 4000 15th Avenue NE, WA98195-1310, Seattle, U.S.A.}

\author{Michael Krisch}
\affiliation{European Synchrotron Radiation Facility, 71, Avenue des Martyrs, F-38000 Grenoble, France}

\date{\today}

\begin{abstract}
We have determined the lattice dynamics of MgSiO$_3$ perovskite (bridgmanite) by a combination of single-crystal inelastic x-ray scattering and \textit{ab initio} calculations. We observe a remarkable agreement between experiment and theory, and provide accurate results for phonon dispersion relations, phonon density of states and the full elasticity tensor. The present work constitutes an important milestone to extend this kind of combined studies to extreme conditions of pressure and temperature, directly relevant for the physics and the chemistry of Earth’s lower mantle.
\end{abstract}



\maketitle

\section{Introduction}
The Earth’s lower mantle, spanning from 670 to 2900 km in depth, represents roughly 60\% of Earth’s total volume. Numerous studies indicate that its chemical composition is likely dominated by a perovskite-structured mineral phase, with MgSiO$_3$ being its main constituent, with Fe and Al as minor elements \cite{andrault_epsl_2001,hummer_am_2012,murakami_nature_2012}. Accordingly, magnesium perovskite is likely the most abundant mineral on the Earth. However, irrespectively of its abundance, the natural occurrence of MgSiO$_3$ perovskite was reported only very recently \cite{tschauner_science_2014}, in the Tenham L6 chondrite, a shocked meteorite, as the mineral is metastable under ambient condition and unpreservable during the exhumation of terrestrial mantle rocks. After the discovery, MgSiO$_3$ has been named bridgmanite \cite{tschauner_science_2014}. 

The determination of the physical properties of bridgmanite is of fundamental geophysical interest. Indeed, while the analysis of seismic records allows the determination of compressional and shear velocities in situ, the establishment of a compositional and mineralogical model for the mantle requires the knowledge of the mineral’s equation of state and its elastic and thermodynamic properties at the relevant pressure (P) and temperature (T) conditions. Structural studies at high pressure and high temperature by X-ray diffraction on samples compressed in a laser-heated diamond anvil cell have been extensively performed \cite{fiquet_pepi_1998, fiquet_grl_2000, ballaran_epsl_2012}  and provide important insight into the pressure-volume (P-V) and pressure-volume-temperature (P-V-T) relationships. The elastic moduli of MgSiO$_3$ perovskite at ambient conditions have been measured on both single-crystal and polycrystalline samples \cite{yeganeh_pepi_1994, sinogeikin_grl_2004}. The aggregate shear modulus has been measured up to 96 GPa \cite{murakami_epsl_2007}. Numerous \textit{ab initio} calculations have been performed as well, with the aim not only to validate the experimental structural studies, but also to predict thermodynamic and elastic properties, phase stability, and effects of anharmonicity via the calculation of the phonon dispersion \cite{parlinski_epb_2000, wentzcovitch_prl_2004, oganov_jcp_2005, carrier_prb_2007, zhang_prl_2014}. Noteworthy, despite this large body of theoretical calculations, there is not a single experimental phonon dispersion study. This is largely explained by the fact that single crystals of sufficiently good quality are only available in very small quantities. This excludes the utilization of inelastic neutron scattering measurements, but it presents a unique opportunity for inelastic x-ray scattering (IXS) with meV energy resolution. In fact, IXS offers the possibility to study samples in diamond anvil cells (DACs), thus enabling combined pressure-temperature studies at Mbar pressure and temperatures up to 1100 K \cite{fiquet_pepi_2004, krisch_Springer_2007, ghose_prl_2006, antonangeli_science_2011, antonangeli_epsl_2012}. 

In the present work we utilize IXS to determine the phonon dispersion of single crystal MgSiO$_3$ along the main symmetry directions at ambient conditions. The experimental results are confronted with state-of-the-art \textit{ab initio} calculations. We benchmark our results by comparing the derived elastic moduli with Brillouin light scattering measurements and the energies of Raman active modes with experimental values available in literature.

\section{Experimental Details}
The bridgmanite crystals were synthesized starting from 85 mol\% MgSiO$_3$ + 15 mol\% Mg$_2$SiO$_4$ mixture as a silicate source. H$_2$O was used as a solvent, and the bulk water content of the sample was about 13 wt\%. Synthesis conditions were P = 24 GPa and T = 1500 $^{\circ}$C, with heating duration of 20 min using a Kawai type 5000 ton multianvil press at Okayama University. Details of the synthesis procedure of bridgmanite can be found elsewhere \cite{shatskiy_am_2007}. The water content is estimated to be about 140(52) wt ppm using polarized IR spectra of thin sections perpendicular to a, b, and c axes of bridgemanite.

Optically transparent samples were characterized by single-crystal X-ray diffraction using an Xcalibur diffractometer (Rigaku-Oxford Diffraction) using Mo K$\alpha$ radiation (MoK$\alpha_1$ 0.70936 \AA, MoK$\alpha_2$ 0.71359 \AA) and a CCD detector. Most of the samples proved to be twins, except one. Diffraction data were collected on this sample (50 $\times$ 50 $\times$ 100 $\mu$m in size) via omega scans (66.57 mm detector-to-sample distance, 25 seconds per frame, 1$^{\circ}$ frame width) up to 37$^{\circ}$ in Bragg theta angle. The crystal structure of MgSiO$_3$ bridgmanite, orthorhombic, \textit{Pnma}, consists of a framework of corner-linked SiO6 octahedra with Mg in an irregular 10-fold coordination. We determined lattice constants a = 4.9436(3) Å, b = 6.9149(8) Å and c = 4.7902(6) Å. Data collected were merged obtaining a discrepancy factor among symmetry related reflections (internal discrepancy factor R$_{int}$), R$_{int}$ = 3.1\%, and typical rocking curve widths were of about 0.4$^{\circ}$. The structural refinement was carried out using the least-squares program SHELXL \cite{sheldrick_ac_2008} in the space group \textit{Pnma} starting from the coordinate of Yagi \textit{et al}\cite{yagi_pcm_1978}. Anisotropic refinement cycles converged to a discrepancy factor R$_1$=2.5\% for 535 Fo$>$4$\sigma$Fo and a goodness of fit of 1.106. Bonds and angles are similar to literature data \cite{yagi_pcm_1978}.
The IXS experiment was carried out at the Inelastic X-ray Scattering Beamline II (ID28) at the European Synchrotron Radiation Facility in Grenoble (France), utilizing the Si (9 9 9) configuration, which provides an overall instrument energy resolution of 3 meV at 17 794 keV. X-ray beam was focused to a spot size at the sample position of 30 (horizontal) by 80 (vertical) $\mu$m$^2$ full-width-half-maximum (FWHM). Direction and size of the momentum transfer were selected by an appropriate choice of the scattering angle and the sample orientation in the horizontal scattering plane. The momentum resolution was typically set to 0.28 nm$^{-1}$ and 0.84 nm$^{-1}$ in the horizontal and vertical plane, respectively. Further details of the experimental set-up can be found elsewhere \cite{krisch_Springer_2007}.

\section{Lattice Dynamics Calculation}
First principle calculations were performed within the generalized gradient approximation to the density functional theory in Perdew-Burke-Ernzerhof parameterization as implemented in the CASTEP code \cite{clark_zkri_2005}. The pseudopotentials were of the norm-conserving type and created using the optimized method of Rappe et al. with single projectors \cite{rappe_prb_1990}. Three reference orbitals were treated as valence states for magnesium and silicon, two for oxygen. The wave functions are expanded in plane waves up to a kinetic cut-off energy of 800 eV at special k-points on a 4 $\times$ 3 $\times$ 4 Monkhorst Pack grid resulting in a convergence for internal forces of $<$10$^{-4}$ eV/Å. The structure optimization was performed using the Broyden-Fletcher-Goldfarb-Shannon method by varying cell parameters and internal coordinates \cite{pfrommer_jcp_1997}. The calculated lattice parameters at zero temperature agree within 1.1\% with the experimental values at 77K \cite{ross_pcm_1989} (see Table S1 in the supplementary information). The dynamical matrices were computed on a 5 $\times$ 3 $\times$ 5 Monkhorst Pack grid employing density functional perturbation theory and were further Fourier interpolated in the cumulate scheme including all image force constants \cite{refson_prb_2006,parlinski_prl_1997,gonze_prb_1994}. The well-converged internal forces yield a maximum error in phonon energies of $<$0.1 meV. Sum rule corrections for acoustic phonons and Born charges were applied with a maximal correction of 3 meV at $\Gamma$. IXS intensities were calculated in first order approximation assuming the validity of both harmonic and adiabatic approximation \cite{krisch_Springer_2007}. Thermodynamic properties are evaluated within the quasi-harmonic approximation.
The elasticity tensor was calculated using a least-square fit as detailed below. A sphere in reciprocal space containing 485 q-points close to $\Gamma$ with $\mid$q$\mid \leq$ 0.02 relative lattice units (r.l.u.) was used for this fit and the dynamical matrices on these q-points were computed using Fourier interpolation. According to the definition of acoustic phonons, the deformation energies can be calculated at each q-point, using the dynamical matrix and the same displacement modulus and polarization for all atomic sites. A given choice of displacement and q-vector determine the deformation tensor. In elasticity theory, the deformation energy can be expressed as a linear combination of the elastic constants for the given deformation tensor. We impose via a least square fit that the two ways of calculating the deformation energy coincide for all the considered q-points and polarizations, and obtain in this way the elasticity tensor.

\section{Results}
\label{sec:results}
\figuremacro{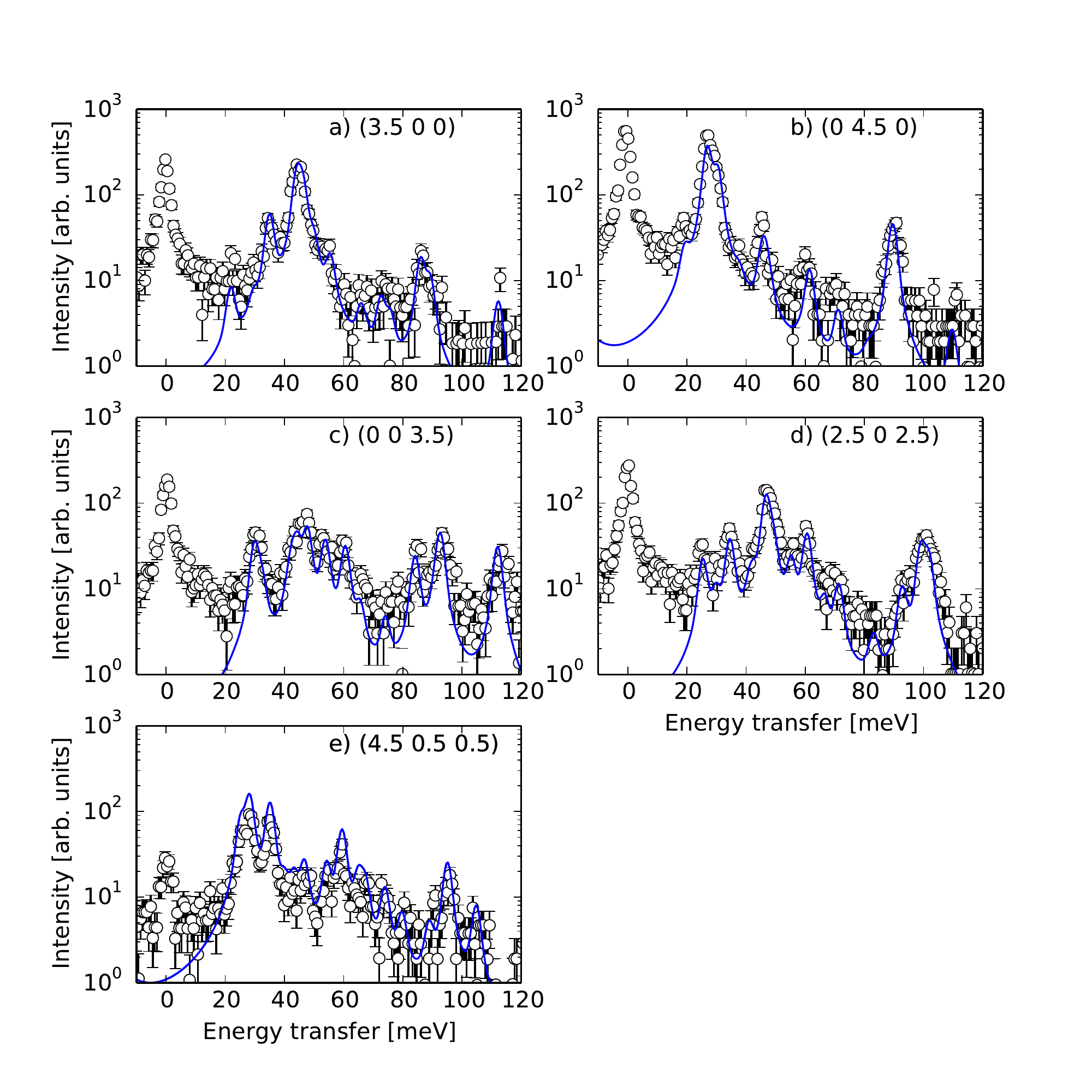}{Experimental (open symbols with corresponding error bars) and calculated (blue lines) IXS spectra of single crystal MgSiO$_3$ at selected reciprocal space points (a) X, b) Y, c) Z d) U and e) R. The \textit{ab initio} calculated intensities were convoluted with the experimental resolution function and the calculated energy transfer scaled by 1.05.}{1.0}

Figure \ref{fig:MgSiO3_ixs_scans} illustrates representative IXS spectra at selected reduced momentum transfers as specified in the legends and figure caption. The experimental spectra are characterized by an elastic line, centered at zero energy transfer, and several inelastic features in the energy range between 20 and 110 meV, corresponding to acoustic and optical phonons. A logarithmic intensity-scale is chosen for better visualization of the high-energy phonon modes, which have a much weaker intensity. Solid blue lines in the individual panels represent the results of the theoretical calculations of the inelastic contribution. The computed energy transfer scale was adjusted by a single factor of 1.05, and, for each q-value, the intensities were scaled by the same factor to obtain best agreement with the experimental results. The contribution at zero energy transfer, due to elastic and quasi-elastic diffuse scattering, is observed in the experiments but is not accounted for by the calculation.

\figuremacro{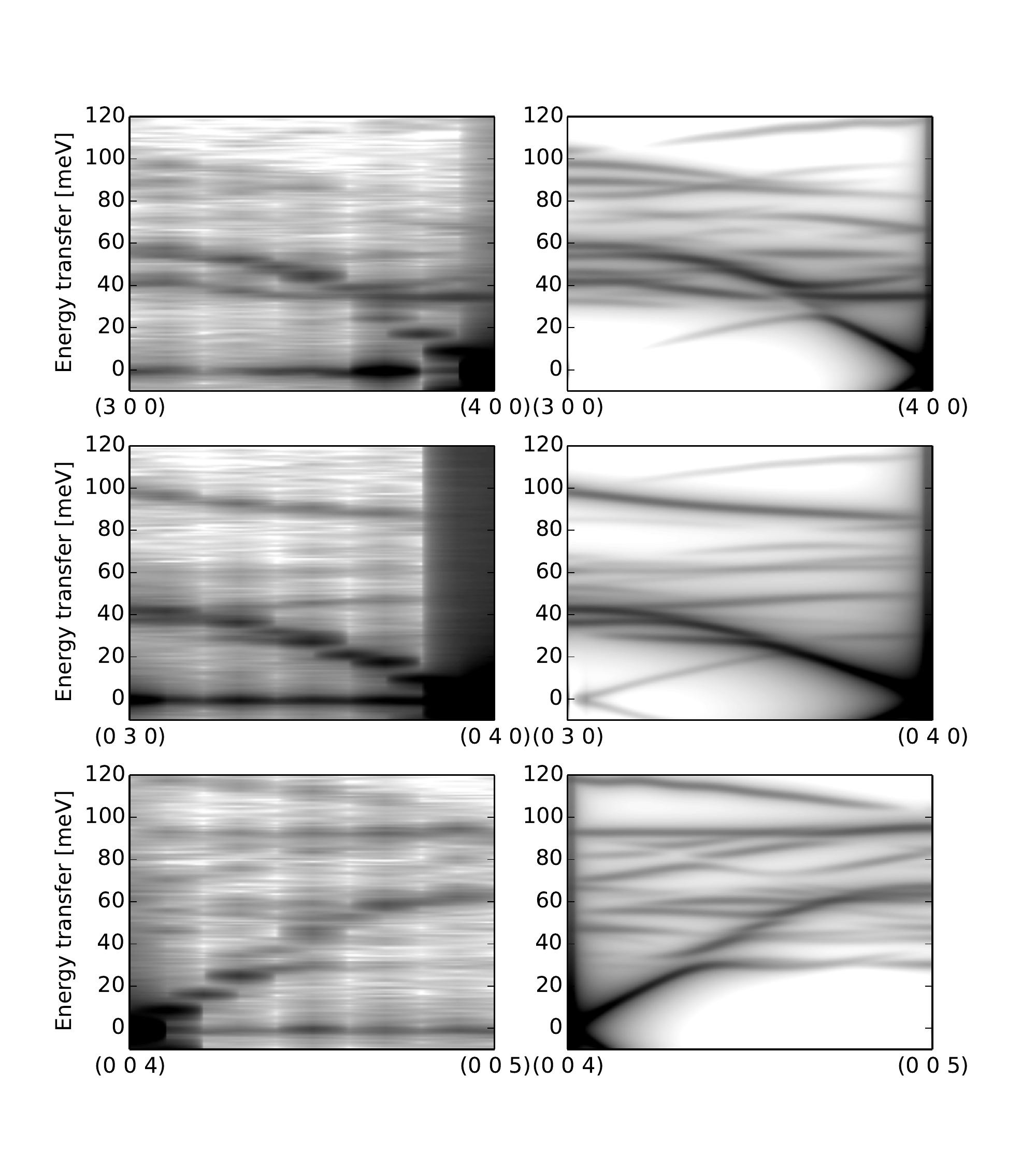}{Experimental (left panels) and calculated (right panels) IXS intensity maps along the indicated directions on a logarithmic intensity scale. The experimental maps consist each of eleven spectra linearly interpolated to 200 Q-points. The calculated intensities are convoluted with the experimental resolution function and the energy transfer is scaled by 1.05.}{1.0}

Figure \ref{fig:MgSiO3_ixs_maps} compares experimental and theoretical IXS intensity maps along three selected principal directions in reciprocal space. These give a more global overview on the complex phonon dispersion and the intensity evolution as a function of momentum transfer Q. Besides a strong contribution from the elastic line close to the (0 4 0) point and some artifacts due to the interpolation routine used for the reconstruction of the IXS maps from individual fixed-Q IXS experimental spectra, we note a very good agreement in terms of phonon energies and intensities. Together with the results displayed in Figure 1 this proves that our calculations not only correctly predict the phonon energies (after a modest overall scaling), but also the phonon eigenvectors, which govern the intensities \cite{bosak_grl_2009}. The slight underestimation of the calculated energies can be mainly attributed to limitations in accuracy of the exchange correlation functional within the general gradient approximation (see Ref. \cite{refson_prb_2006} for a detailed discussion).

\figuremacro{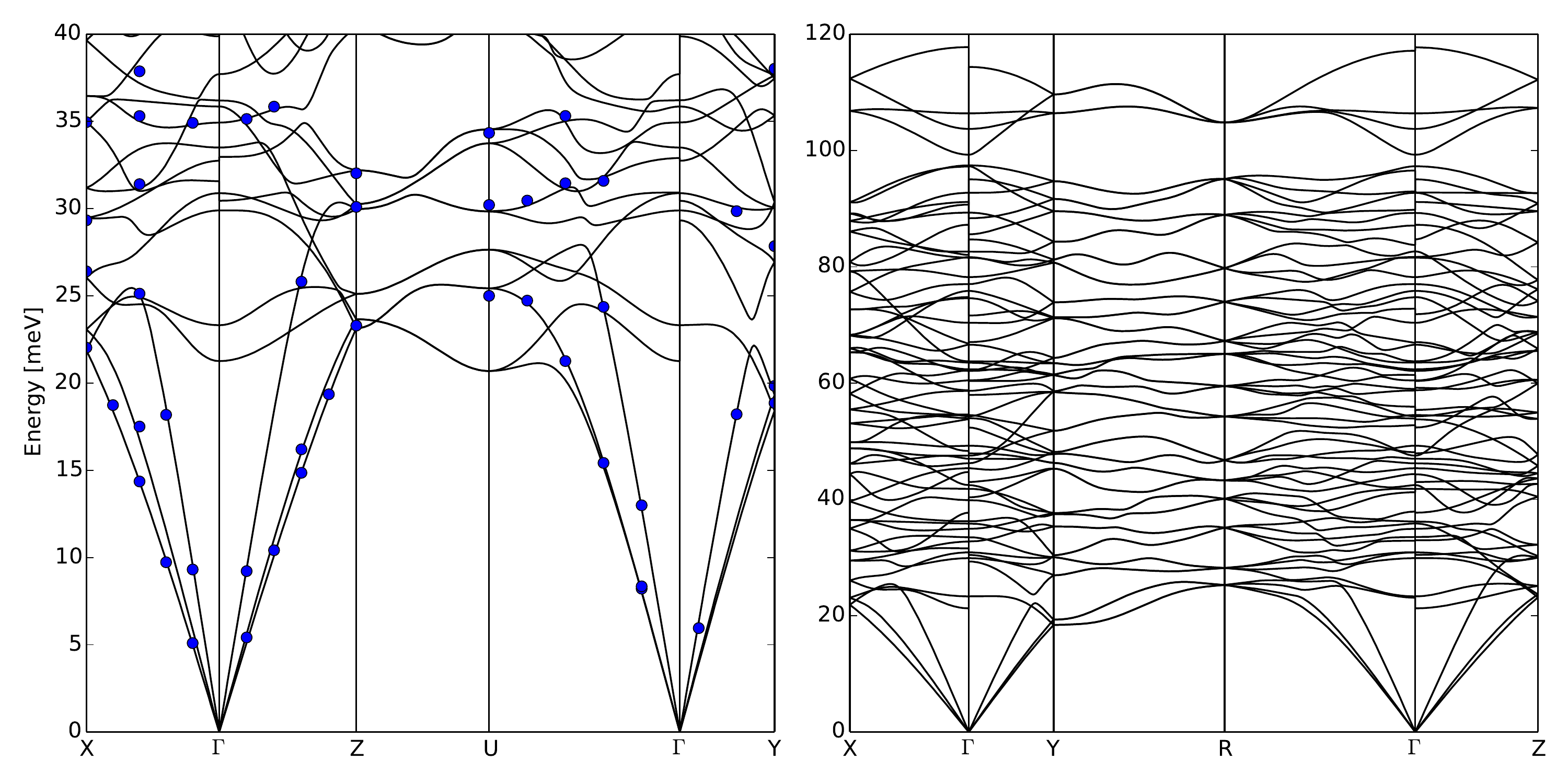}{Calculated phonon dispersion relations (solid lines) along selected high symmetry directions. The left panel is a zoom of the low energy part; the right panel covers the full energy range. Experimentally determined phonon energies are plotted as blue dots, the calculated energies are scaled by 1.05.}{1.0}

\begin{table}[tbp]
\caption{\label{tab:Raman}Phonon energies of the Raman active modes MgSiO$_3$ in meV. The \textit{ab initio} calculated values are scaled by 1.05 and compared to experimental values at 290K \cite{gillet_pepi_2000}. The intensities are classified by very strong (vs), strong (s), weak (w) and very weak (vw).} 

\begin{tabular}{c c c c}
\toprule
\multicolumn{2}{c}{calculation} & \multicolumn{2}{c}{experiment} \\
\midrule
29.9 & s & 31.0 & s \\
30.9 & s & 31.6 & s \\
34.9 & s & 35.0 & s \\
35.9 & vw & 35.7 &vw \\
36.2 & vw & - \\
41.8 & s & 41.8 & s \\
42.5 & w & 42.4 & w \\
46.2 & s & 45.8 &s \\
47.0 &vs & 47.4 & vs \\
49.2 & s & 48.6 & s \\
54.2 & w & - \\
54.5 & vw & - \\
62.1 & s & 62.9 & s \\
62.4 & vs & 62.1 & vs \\
63.5 & s & - \\
63.7 & s & 67.7 w \\
66.6 & vs & 67.1 &s \\
77.0 & w & - \\
78.2 & w & - \\
81.7 & s & - \\
82.6 & s & 82.6 & s \\
99.3 & vs & - \\
103.7 & vw & - \\
106.4 & w & - \\
\bottomrule

\end{tabular}
\end{table}

In Figure 3 we display the MgSiO$_3$ phonon dispersion along selected high symmetry directions. For the low-energy part we superimpose the experimentally determined phonon energies. These were extracted by fitting a set of Lorentzian functions convoluted with the experimental resolution function to the IXS spectra, utilizing a standard $\chi^2$ minimization routine. The agreement with the calculation is excellent. Inspection of the complete phonon dispersion - which extends up to 120 eV - reveals that the correct assignment of an experimentally observed phonon mode becomes increasingly more difficult due to the high density of phonon branches. Nonetheless, the already mentioned remarkable agreement between experiment and theory (after a mere rescaling) makes us confident that the calculation correctly describes the lattice dynamics of MgSiO$_3$ even at the higher energies not experimentally covered. The phonon dispersion of the optical branches is rather complex and shows pronounced splitting of longitudinal optical and transvers optical branches. The splitting is due to large Born effective charges. MgSiO$_3$ exhibits 24 Raman active modes of which 14 have been determined by experiments \cite{gillet_pepi_2000}. The calculated energies are given and compared to experiments in Table \ref{tab:Raman}. The values were assigned by careful comparison of calculated and measured intensities. The mode at 99.3 meV should be visible in experiment, but is beyond the measured energy transfers of reported measurements. The agreement between experiment and calculation is very good with the only exception of the mode with a calculated energy of 63.7 meV. 

\figuremacro{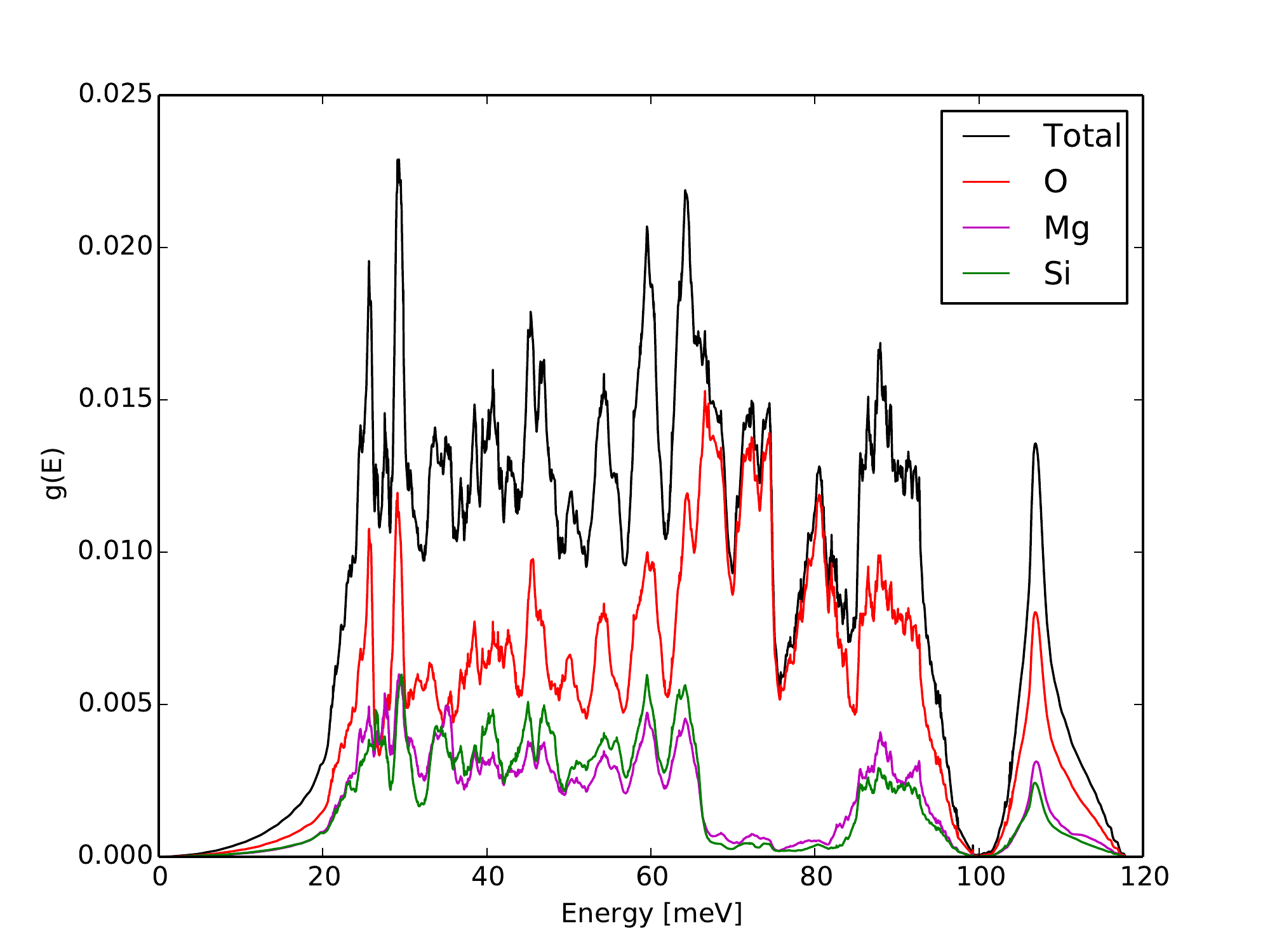}{Calculated phonon density of states of MgSiO$_3$. The partial density of states are given per species. The energy is scaled by 1.05.}{1.0}

The so-validated theoretical results were then used to derive the partial and total phonon density-of-states (DOS), shown in Figure \ref{fig:MgSiO3_dos}. The partial contributions per species indicate that the vibration of oxygen atoms dominates the first two low energy peaks in the DOS. Between 67 and 82 meV all modes are almost exclusively due to oxygen. We further note a gap between 100 and 102 meV. The total DOS allows the derivation of important thermodynamic properties via integral equations \cite{jones_wiley_1973}. Some relevant thermodynamic quantities calculated within the quasi-harmonic approximation for room temperatures can be found in the supplementary information (see Table S2).

\begin{table}[tbp]
\caption{\label{tab:elastic}Elastic moduli in GPa derived from the present \textit{ab initio} calculations, by fitting the elasticity tensor to acoustic phonons at low q (DFPT). Experimental values from BLS are reported for comparison \cite{sinogeikin_grl_2004}.} 

\begin{tabular}{ c c c c c c c c c c }
\toprule
 & c$_{11}$ & c$_{22}$ & c$_{33}$ & c$_{44}$ & c$_{55}$ & c$_{66}$ & c$_{12}$ & c$_{13}$ & c$_{23}$  \\
\midrule
DFPT & 511(2) & 426(2) & 494(2) & 176(1) & 151(1) & 193(1) & 149(1) & 109(1) & 	137(1)  \\
BLS & 528 & 456 & 481 & 182 & 147  & 200 & 146 & 125 & 139  \\
\bottomrule
\end{tabular}
\end{table}

The full elasticity tensor is given in Table \ref{tab:elastic}. The elastic moduli were determined using a least square fit to acoustic phonons without the need of symmetry braking displacements. For comparison we also employed the traditional method exploiting the stress-strain relation from 18 distorted structures, which leads to comparable values. The calculated values are in good agreement with literature values determined by Brillouin light scattering (BLS) \cite{sinogeikin_grl_2004}.

The applied method provides a useful tool to derive the elasticity tensor directly from the lattice dynamics. It profits from the acoustic sum rule corrections, which overcome numerical issues due to the finite grid sampling of reciprocal space and finite number of plane waves. As further advantage, in contrast to stress-strain calculations using distorted structures, all structural symmetry operations can be exploited to reduce the number of k-points for the Brillouin zone sampling. The calculations are thus much easier to converge.

\section{SUMMARY AND PERSPECTIVES}
Our experimental study combined with parameter-free model calculations provides the complete lattice dynamics description of MgSiO$_3$-brigmanite, the main constituent of the Earth’s lower mantle and one of the most important Earth’s minerals. More in general, our work clearly demonstrates that the phonon related properties of many geophysically relevant materials can nowadays be obtained without the need to record the phonon branches one-by-one along all the high symmetry directions. We have shown that the determination of a representative sub-set of experimental data is sufficient for the reliable determination of the full lattice dynamics. Consequently, very important thermodynamic and elastic properties can be derived with high accuracy. In particular, the knowledge of the full phonon dispersion is crucial for validating any physical quantity related to the lattice dynamics. Furthermore, when extended to high pressure and high temperature conditions, it allows the validation of the phase stability proposed on the basis of energy minimization. It is worth noting that the availability of a single crystal, while beneficial, is not mandatory, and the very same concept can be applied to polycrystalline materials as was demonstrated in the case of stishovite \cite{bosak_grl_2009}. 
The here-reported study at ambient conditions marks the starting point for a comprehensive investigation of the stable phases in Earth’s mantle, including Mg$_2$SiO$_4$ (forsterite, spinel, $\beta$-phase), MgSiO$_3$ (orthopyroxene, ilmenite, perovskite), and SiO$_2$ (stishovite, seifertite) at relevant pressure and temperature conditions of the Earth’s lower mantle. Such a study is technically feasible as the x-ray beam at ID28 can be focused down to 17$\times$ 10 $\mu$m$^2$ with a flux reduction of only $\approx$ 25-30\% compared to the optical configuration used in the present experiment. This allows working efficiently with samples of lateral dimensions of few tens of microns. Once considering samples down to a thickness of 15 – 20 $\mu$m, compatible with experiments in a diamond anvil cell (DAC) up to Mbar pressures, the expected loss in the IXS signal is 5-7 with respect to the present experiment. This would lead to typical acquisition times per spectrum of about 4h, and a complete set of data in a few days. It is worth noting that modern IXS spectrometers are equipped with several crystal analyzers (nine in the case of ID28); thus IXS spectra at several different momentum transfers are recorded simultaneously.

\section*{Acknowledgment}
We are grateful to Artem Oganov and Guangrui Qian for preliminary phonon calculations. Beamtime was granted on beamline ID28 at the ESRF in the frame of proposal HS-4807. This paper is devoted to Subrata Ghose, a friend and colleague. He was a visionary mineralogist and a great man.

\bibliographystyle{apsrev4-1_BW}
\bibliography{MgSiO3_refs}

\clearpage

\section*{SUPPLEMENTARY INFORMATION}

\setcounter{equation}{0}
\setcounter{figure}{0}
\setcounter{table}{0}
\renewcommand{\theequation}{S\arabic{equation}}
\renewcommand{\thefigure}{S\arabic{figure}}
\renewcommand{\thetable}{S\arabic{table}}

\begin{table}[hp]
\caption{Lattice constants of MgSiO$_3$ perovskite (space groupe \textit{Pnma}).} 
\label{tab:lattice}
\begin{tabular}{ c c c c }
\toprule
 & Calculation [Å] & Experiment (77K) [Å] \cite{ross_pcm_1989}  & Our experiment (297 K) \\
\midrule
a & 4.979 & 4.930  & 4.9436(4) \\
b & 6.973 & 6.896 & 6.9149(8) \\
c & 4.828 & 4.773 & 4.7902(6) \\
\bottomrule
\end{tabular}
\end{table}

\begin{table}[hp]
\caption{Calculated thermodynamic properties per unit cell at 298K. Vibrational contribution to the total energy E and free energy F; Entropy S and heat capacity C$_v$.} 
\label{tab:thermo}
\begin{tabular}{ c c c c }
\toprule
E (eV) & F(eV) & S (J/mol/K) & C$_v$ (J/mol/K) \\
0.476 & -0.252 & 244.4 & 332.9 \\
\bottomrule
\end{tabular}
\end{table}

\end{document}